\documentclass[12pt, a4paper]{article}
\usepackage{epsf}
\usepackage{cite}
\usepackage{amsmath,amssymb}
\input{colordvi.tex}
\usepackage[dvips]{graphicx}
\usepackage{comment}
\usepackage[height=22.5cm,width=16.5cm,centering
]{geometry}

\begin{document}

\newcommand{\lsim}{\stackrel{<}{_\sim}}
\newcommand{\gsim}{\stackrel{>}{_\sim}}
\newcommand{\rem}[1]{{$\spadesuit$\bf #1$\spadesuit$}}

\renewcommand{\theequation}{\thesection.\arabic{equation}}
\renewcommand{\thefootnote}{\fnsymbol{footnote}}
\setcounter{footnote}{0}

%\parindent = 0pt
%%%%%%%%%%%%%%%%%%%%%%%%%%%%%%%%%%%%%%%%%%%%%%%%%%

%%%%%%%%%%%%%%%%%%%%%%%%%%%%%%%%%%%%%%%%%%%%%%%%%%
\def\thefootnote{\fnsymbol{footnote}}
\def\a{\alpha}
\def\b{\beta}
\def\c{\varepsilon}
\def\d{\delta}
\def\e{\epsilon}
\def\f{\phi}
\def\g{\gamma}
\def\h{\theta}
\def\k{\kappa}
\def\l{\lambda}
\def\m{\mu}
\def\n{\nu}
\def\p{\psi}
\def\q{\partial}
\def\r{\rho}
\def\s{\sigma}
\def\t{\tau}
\def\u{\upsilon}
\def\v{\varphi}
\def\w{\omega}
\def\x{\xi}
\def\y{\eta}
\def\z{\zeta}
\def\D{\Delta}
\def\G{\Gamma}
\def\H{\Theta}
\def\L{\Lambda}
\def\F{\Phi}
\def\P{\Psi}
\def\S{\Sigma}
\def\me{\mathrm e}

\def\o{\over}
\def\beq{\begin{eqnarray}}
\def\eeq{\end{eqnarray}}
\newcommand{\vev}[1]{ \left\langle {#1} \right\rangle }
\newcommand{\bra}[1]{ \langle {#1} | }
\newcommand{\ket}[1]{ | {#1} \rangle }
\newcommand{\bs}[1]{ {\boldsymbol {#1}} }
\newcommand{\mc}[1]{ {\mathcal {#1}} }
\newcommand{\mb}[1]{ {\mathbb {#1}} }
\newcommand{\EV}{ {\rm eV} }
\newcommand{\KEV}{ {\rm keV} }
\newcommand{\MEV}{ {\rm MeV} }
\newcommand{\GEV}{ {\rm GeV} }
\newcommand{\TEV}{ {\rm TeV} }
\def\diag{\mathop{\rm diag}\nolimits}
\def\Spin{\mathop{\rm Spin}}
\def\SO{\mathop{\rm SO}}
\def\O{\mathop{\rm O}}
\def\SU{\mathop{\rm SU}}
\def\U{\mathop{\rm U}}
\def\Sp{\mathop{\rm Sp}}
\def\SL{\mathop{\rm SL}}
\def\tr{\mathop{\rm tr}}
\def\sp{\;\;}

\def\IJMP{Int.~J.~Mod.~Phys. }
\def\MPL{Mod.~Phys.~Lett. }
\def\NP{Nucl.~Phys. }
\def\PL{Phys.~Lett. }
\def\PR{Phys.~Rev. }
\def\PRL{Phys.~Rev.~Lett. }
\def\PTP{Prog.~Theor.~Phys. }
\def\ZP{Z.~Phys. }
%%%%%%%%%%%%%%%%%%%%%%%%%%%%%%%%%%%%%%%%%%%%%%%%%%

%%%%%%%%%%%%%%%%%%%%%%%%%%%%%%%%%%%%%%%%%%%%%%%%%%
\begin{titlepage}

\begin{center}

\hfill UT-15-19\\
\hfill IPMU-15-0071\\
\hfill May, 2015\\

\vskip .75in

{\Large \bf 
On adiabatic invariant in generalized Galileon theories
}

\vskip .75in

{\large Yohei Ema$^a$, Ryusuke Jinno$^a$, Kyohei Mukaida$^b$ and Kazunori Nakayama$^{a,b}$}

\vskip 0.25in

\begin{tabular}{ll}
$^{a}$&\!\! {\em Department of Physics, Faculty of Science, }\\
& {\em The University of Tokyo,  Bunkyo-ku, Tokyo 133-0033, Japan}\\[.3em]
$^{b}$ &\!\! {\em Kavli IPMU (WPI), UTIAS,}\\
&{\em The University of Tokyo,  Kashiwa, Chiba 277-8583, Japan}
\end{tabular}

\end{center}
\vskip .5in

\begin{abstract}
We consider background dynamics of generalized Galileon theories in the context of inflation, 
where gravity and inflaton are non-minimally coupled to each other.
In the inflaton oscillation regime, the Hubble parameter and energy density oscillate violently in many cases, in contrast to 
the Einstein gravity with minimally coupled inflaton.
However, we find that there is an adiabatic invariant in the inflaton oscillation regime in any generalized Galileon theory.
This adiabatic invariant is useful in estimating the expansion law of the universe
and also the particle production rate due to the oscillation of the Hubble parameter.
\end{abstract}

\end{titlepage}

\tableofcontents

\renewcommand{\thepage}{\arabic{page}}
\setcounter{page}{1}
\renewcommand{\thefootnote}{\#\arabic{footnote}}
\setcounter{footnote}{0}
%%%%%%%%%%%%%%%%%%%%%%%%%%%%%%%%%%%%%%%%%%%%%%%%%%

\newpage

%%%%%%%%%%%%%%%%%%%%%%%%%%%%%%%%%%%%%%%%%%%%%%%%%%
\section{Introduction}
\label{sec:introduction}
\setcounter{equation}{0}
%%%%%%%%%%%%%%%%%%%%%%%%%%%%%%%%%%%%%%%%%%%%%%%%%%

Inflation~\cite{Guth:1980zm} is now widely accepted as a successful way of producing the flat and isotropic universe 
as well as small density perturbations as seeds for galaxies.
It is triggered by the potential energy of a scalar field called inflaton, 
which slowly rolls down its potential during inflation~\cite{Linde:1981mu}.
For successful reheating,
inflaton inevitably couples to light particles, and
this coupling may induce sizable radiative corrections to its effective potential.
Even if this is not the case, there might be a self interaction term of the inflaton.
As a result, it might contradict with the present observational data
by producing too large density perturbations~\cite{Ade:2015lrj}.
Among several approaches to this issue, one way is to generalize the interaction between inflaton and gravity
to flatten its effective potential.
However, the coupling of inflaton to gravity in general produces higher derivatives in their equations of motion.
Since it can induce other degrees of freedom which might lead to ghost instabilities,
it would be desirable to construct models without introducing new degrees of freedom.
There is a class of theories with a scalar field and gravity, called
generalized Galileons\cite{Deffayet:2011gz}, which do not suffer from such higher derivatives.
Efforts have been made to study general setups for using these theories in inflation 
(generalized G-inflation\cite{Kobayashi:2011nu}), and
such theories are under close investigation especially
after the discovery of the Higgs boson at the LHC\cite{Aad:2012tfa,Chatrchyan:2012ufa} and 
after the results from Planck satellite\cite{Ade:2015lrj}, in the context of identifying the inflaton as 
the Higgs field\cite{Futamase:1987ua,CervantesCota:1995tz,Bezrukov:2007ep,Germani:2010gm,Takahashi:2010ky}.

In this paper, we study the oscillation regime of the generalized Galileon theories in the context of inflation,
because it is necessarily in order to understand the full reheating dynamics and construct a consistent cosmological scenario.
In fact, it has recently been pointed out that in some models of the generalized G-inflation
the Hubble parameter oscillates violently during the inflaton oscillation regime\cite{Jinno:2013fka},
which makes the analysis of expansion law of the universe non-trivial and also it may lead to Laplacian instability~\cite{Ema:2015oaa}.
The knowledge of the expansion law of the universe in these models is also necessarily
to make a precise prediction for the spectral index of the density perturbation, for example.
For these purposes, we construct a useful quantity, which we call an adiabatic invariant,  
in analyzing the background dynamics.
The expansion law of the universe during inflaton oscillation regime is easily derived by using this invariant.
It is also useful to extract the oscillation part of the Hubble parameter, 
which helps us estimate the particle production rate via gravitational effect.

The organization of this paper is as follows.
In Sec.~\ref{sec:general}, we explain the procedure to obtain the adiabatic invariant, 
to express the oscillation of the Hubble parameter in terms of the inflaton, and
to obtain the expansion law of the universe. 
In Sec.~\ref{sec:examples}, we illustrate this procedure using some examples.
The last section is devoted to conclusion.

%%%%%%%%%%%%%%%%%%%%%%%%%%%%%%%%%%%%%%%%%%%%%%%%%%
\section{General discussion}
\label{sec:general}
\setcounter{equation}{0}
%%%%%%%%%%%%%%%%%%%%%%%%%%%%%%%%%%%%%%%%%%%%%%%%%%

We consider the following action throughout this paper:
\begin{align}
S
&= S_G + S_M,
\label{eq:action}
\end{align}
where $S_G$ is the action for the gravity and the inflaton $\phi$, including non-minimal couplings between them,
and $S_M$ is the action for the matter field.\footnote{
Though we focus on the oscillation regime of single-field inflation, 
generalization of the adiabatic invariant to multi-field case is trivial 
since, as we will see later, the necessary equation is only Eq.~(\ref{eq:eoma}).
}
We assume that $S_G$ is given by the Lagrangian for generalized Galileon theories\cite{Deffayet:2011gz}:
\begin{align}
S_G
&= \int d^4x \sqrt{-g} \; \sum_{i=2}^5 {\mathcal L}_i,
\end{align}
where
\begin{align}
{\mathcal L}_2 
&= G_2(\phi, X), 
\label{eq:G_2}\\
{\mathcal L}_3 
&= -G_3(\phi, X)\Box \phi, 
\label{eq:G_3} \\
{\mathcal L}_4 
&= G_4(\phi, X)R + G_{4X}\left[ (\Box \phi)^2 - (\nabla_\mu \nabla_\nu \phi)^2 \right], 
\label{eq:G_4} \\
{\mathcal L}_5 
&=G_5(\phi, X)G^{\mu \nu}\nabla_\mu \nabla_\nu \phi
- \frac{G_{5X}}{6}\left[ (\Box \phi)^3 - 3(\Box \phi)(\nabla_\mu \nabla_\nu \phi)^2 + 2(\nabla_\mu \nabla_\nu \phi)^3 \right].
\label{eq:G_5}
\end{align}
Here $G_i$'s are general functions of $\phi$ and $X \equiv -g^{\mu \nu}\nabla_\mu \phi \nabla_\nu \phi / 2$, 
and the subscript $X$ denotes the derivative respect to $X$.
Moreover, 
$(\nabla_\mu \nabla_\nu \phi)^2$ and $(\nabla_\mu \nabla_\nu \phi)^3$ are understood as 
$\nabla_\mu \nabla_\nu \phi \nabla^\mu \nabla^\nu \phi$ and 
$\nabla_\mu \nabla_\nu \phi \nabla^\nu \nabla^\rho \phi \nabla_\rho \nabla^\mu \phi$, respectively.
We take the background part of the metric to be 
the Friedmann-Lema\^itre-Robertson-Walker (FLRW) one with negligible curvature,
\begin{align}
ds^2
&= -dt^2 + a(t)^2 \delta_{ij} dx^idx^j,
\end{align}
where $a$ is the scale factor. We also assume that the background part of the inflaton is homogeneous, {\it i.e.}~$\phi = \phi(t)$.
In such a case, the background parts of the actions for Eqs.~\eqref{eq:G_2}-\eqref{eq:G_5} are respectively given by
\begin{align}
\mathcal{L}_{2} &= G_{2}, 
\label{eq:L_2} \\
\mathcal{L}_{3} &= -G_{3\phi}\dot{\phi}^{2} - G_{3X}\dot{\phi}^{2}\ddot{\phi}, \\
\mathcal{L}_{4} &= -6H^{2}G_{4}-6HG_{4\phi}\dot{\phi} + 6H^{2}G_{4X}\dot{\phi}^{2}, \\
\mathcal{L}_{5} &= -3H^{2}G_{5\phi}\dot{\phi}^{2} + H^{3}G_{5X}\dot{\phi}^{3},
\label{eq:L_5}
\end{align}
after integration by parts\footnote{
It is known that some of these models have the ghost instability
after inflation, and thus it is even non-trivial whether the inflaton can oscillate around the minimum of the potential
after inflation~\cite{Ohashi:2012wf}. 
The terms which potentially lead to such ghost instabilities are understood as 
those with $\dot{\phi}^{2}$ in Eqs.~\eqref{eq:L_2}-\eqref{eq:L_5}
whose coefficients are not positive definite.
In this paper, we simply neglect such models 
because they may not be connected to the standard big-bang cosmology.
}, where $H=\dot{a}/a$ is the Hubble parameter, the dot denotes 
the derivative with respect to $t$ and the subscripts of $X$ and $\phi$ are the derivative with respect to those quantities. 
Note that all terms except for $\mathcal{L}_{3}$ depend only on $H$, $\dot{\phi}$ and $\phi$. 
In fact, the $\ddot{\phi}$ dependence of $\mathcal{L}_{3}$ can also be eliminated by integration by parts.\footnote{
This can be done for $G_{3}$ polynomial for $\phi$ and $X$ as
\begin{align}
\phi^m X^n \ddot{\phi}\dot{\phi}^2
&= \frac{1}{2n+3}\left( \frac{1}{2} \right)^n \phi^m \left( \dot{\phi}^{2n+3} \right)^\cdot
\rightarrow -\frac{1}{2n+3}\left( \frac{1}{2} \right)^n 
\left[ 3H \phi^m \dot{\phi}^{2n+3} 
+ m \phi^{m-1} \dot{\phi}^{2n+4} \right].
\end{align}
}
Thus, we take the background part of $S_G$ generally as follows:
\begin{align}
S_G
&= \int d^4x \; a^3 {\mathcal L}(H,\dot{\phi},\phi).
\label{eq:gravphiaction}
\end{align}
In this paper, we simply neglect the back reaction to the inflaton due to particle production.\footnote{
It is known that some of these models have a Laplacian instability coming from the negative sound speed squared 
for the scalar perturbation\cite{Ohashi:2012wf,Ema:2015oaa},
and such an instability may lead to a rapid decay of the inflaton condensation.
Anyway, in order to determine whether there is a Laplacian instability or not, we must first solve the background dynamics.
}
We also assume that ${\mathcal L}$ is a sum of polynomials with respect to its arguments.

The background equations of motion for the system (\ref{eq:gravphiaction}) is given by
\begin{align}
&({\mathcal L}_{\dot{\phi}})^\cdot + 3H{\mathcal L}_{\dot{\phi}} - {\mathcal L}_\phi
=0,
\label{eq:eomphi}\\
&\mathcal L - \dot{\phi}{\mathcal L}_{\dot{\phi}} - H{\mathcal L}_H
=0,
\label{eq:eomconst}\\
&({\mathcal L}_H)^\cdot + 3H{\mathcal L}_H - 3{\mathcal L}
=0,
\label{eq:eoma}
\end{align}
where the subscript denotes the derivative with respect to that quantity.
The first equation is $\phi$'s equation of motion and the second one is the Friedmann equation. 
%which can be derived by introducing the lapse function $N$ as $dt \to N \; dt$ and take a variation with respect to it.
In addition, the last equation is obtained by taking functional derivative with respect to the scale factor. 
Note that, as we will see below, the last equation is useful in analyzing the inflaton oscillation regime although
it is redundant.

In this paper, we focus on the inflaton oscillation regime.
We define $m_{\rm{eff}}$ as a typical frequency of the inflaton oscillation in the following discussion.
In general, $m_{\rm{eff}}$ may be estimated as $m_{\rm{eff}} \sim |\dot{\phi}/\phi|$.
In the inflaton oscillation regime, the inequality $m_{\rm{eff}} > H$ is satisfied.
As we will explicitly see in the next section, the Hubble parameter and the energy density of the inflaton
violently oscillate with time in some models, {\it i.e.}~$\dot{H} = {\mathcal O}(m_{\rm eff}H)$.
%In such a case, it is inadequate to evaluate the cosmic expansion law by taking the oscillation average
%with respect to them. 
%Instead, as we will show below, 
However, there is a quantity which 
evolves only with a time scale of $H^{-1}$, not $m_{\rm{eff}}^{-1}$, for the general Lagrangian~\eqref{eq:gravphiaction}.
Such a quantity, which we call an ``adiabatic invariant," can be used to evaluate the cosmic expansion law
in the oscillation regime. In addition, it is also helpful in estimating the amount of particle production via gravitational effects.

\begin{comment}
Such dangerous features may be seen from Eqs.~\eqref{eq:L_2}-\eqref{eq:L_5} as follows.
If the coefficient of $\dot{\phi}^{2}$ is not positive definite, it may lead to the ghost instability of $\phi$
during the inflaton oscillation regime. From Eqs.~\eqref{eq:L_2}-\eqref{eq:L_5}, 
we can see that the $X$-dependent parts of $G_{3}$ and $G_{5}$ may play that role. The $\phi$-dependent parts of
$G_{3}$ and $G_{5}$ may also be dangerous if they have even powers of $\phi$.
In addition, the sign of $G_{4X}$ should be positive to avoid the ghost instability.
\rem{RJ: No mention to $G_{4\phi}$?}
\rem{RJ: Any term in Eqs.~\eqref{eq:L_2}-\eqref{eq:L_5} seems dangerous in general}
In this paper, we simply neglect models which have the ghost instability 
because they cannot naturally be connected to the standard big-bang cosmology.
\end{comment}

%%%%%%%%%%%%%%%%%%%%%%%%%%%%%%%%%%%%%%%%%%%%%%%%%%
\subsection{Adiabatic invariant}
%%%%%%%%%%%%%%%%%%%%%%%%%%%%%%%%%%%%%%%%%%%%%%%%%%

%In this subsection, we explicitly construct an adiabatic invariant other than the scale factor 
%for the general Lagrangian~\eqref{eq:gravphiaction}. 
We define an adiabatic invariant $Q$ by the following equality:
\begin{align}
\dot{Q} = {\mathcal O}(HQ),
\end{align}
which implies that $Q$ evolves only with the time scale of $H^{-1}$.
%Note that the existence of such a quantity is nontrivial\footnote{
%Apart from the scale factor $a$, since it is an adiabatic invariant by definition.
%}
%since we consider the oscillation regime of the inflaton, 
%when the any time derivative which picks up the timescale of $\phi$ oscillation gives $\dot{Q} = {\mathcal O}(m_\phi Q)$.\footnote{
%When the inflaton is oscillating at the minimum of non-quadratic potential,
%one should replace $m_\phi$ with $m_{\rm eff} \equiv \sqrt{|V_\phi / \phi|}$.
%Also, when the inflaton does not have a canonical kinetic term, one should replace $m_\phi$ with
%$m_{\rm eff}$, the mass of the canonically normalized inflaton.
%}
From the Friedmann equation (\ref{eq:eomconst}), we can see that $\mathcal{L} \sim H\mathcal{L}_{H}$ is satisfied.\footnote{
In fact, if $\mathcal{L} \sim H\mathcal{L}_{H}$ does not hold,
${\mathcal L}$ and $\dot{\phi}{\mathcal L}_{\dot{\phi}}$ must cancel almost exactly 
with each other. This occurs only when ${\mathcal L}$ is linear in $\dot{\phi}$, and we do not consider such a case here.
}. Then, from Eq.~(\ref{eq:eoma}) we obtain 
\begin{align}
({\mathcal L}_H)^\cdot
&= {\mathcal O}(H {\mathcal L}_H).
\end{align}
Thus we have proven that ${\mathcal L}_H$ is an adiabatic invariant.
This invariant can also be expressed by $\dot{\phi}$ and $\phi$, 
by eliminating $H$ using Eq.~(\ref{eq:eomconst}).\footnote{
Note that the scale factor $a$ is also an adiabatic invariant by definition.
}

Here we define
\begin{align}
J
&\equiv -\frac{{\mathcal L}_H}{6M_P^2},
\end{align}
and
\begin{align}
\rho_J
&\equiv 3M_P^2 J^2.
\end{align}
Both $J$ and $\rho_{J}$ are adiabatic invariants. Note that they coincide respectively 
with the Hubble parameter and the energy density in the case of the minimal Einstein gravity.
In this sense, $J$ and $\rho_{J}$ may be interpreted as generalization of the Hubble parameter
and the energy density to the case of the generalized Galileon theories, respectively.

%%%%%%%%%%%%%%%%%%%%%%%%%%%%%%%%%%%%%%%%%%%%%%%%%%
\subsection{Evaluation of the cosmic expansion law}
%%%%%%%%%%%%%%%%%%%%%%%%%%%%%%%%%%%%%%%%%%%%%%%%%%
In this subsection, we schematically describe the procedure to evaluate the cosmic expansion law by using 
the adiabatic invariant that we have derived just above. Some specific examples of this procedure will be given in the next section.

In order to obtain the expansion law, we first evaluate the scale factor dependence of ${\mathcal L}_H$.
Let us start with Eq.~\eqref{eq:eoma}. We have to express the oscillation average of $\mathcal{L}$ by that of $H\mathcal{L}_{H}$.
The oscillation average of Eq.~(\ref{eq:eomconst}) gives
\begin{align}
\vev{{\mathcal L}}
- \vev{\dot{\phi}{\mathcal L}_{\dot{\phi}}}
- \vev{H{\mathcal L}_H}
&= 0,
\label{eq:eomconstave}
\end{align}
where the angular bracket denotes an oscillation average which smears out the oscillation with the frequency of 
the order of $\sim m_{\rm{eff}}$. In addition, by multiplying Eq.~(\ref{eq:eomphi}) with $\phi$ and taking an oscillation average,
we obtain the Virial theorem:
\begin{align}
\vev{\dot{\phi}{\mathcal L}_{\dot{\phi}}}
+ \vev{\phi{\mathcal L}_\phi}
&\simeq 0,
\label{eq:virial}
\end{align}
where we have taken only leading terms with respect to $\mathcal{O}(H/m_{\rm{eff}})$.
%Note that we have neglected the oscillation average of total derivatives and  
%$\vev{H\phi{\mathcal L}_{\dot{\phi}}}$, since the latter is negligible compared to $\vev{\dot{\phi}{\mathcal L}_{\dot{\phi}}}$ 
%in Eq.~(\ref{eq:virial}).
Here, note that usually $\mathcal{L}$ can be decomposed into at most three
terms which are proportional to $\dot{\phi}\mathcal{L}_{\dot{\phi}}$, $\phi\mathcal{L}_{\phi}$ 
and $H\mathcal{L}_{H}$,
if one neglects small terms in each regime under consideration.
See also examples in Sec.~\ref{sec:examples}.
Thus, from Eqs.~\eqref{eq:eomconstave} and \eqref{eq:virial} together with such a decomposition, 
one can express $\vev{{\mathcal L}}$ in terms of $\vev{H{\mathcal L}_H}$. Then, Eq.~(\ref{eq:eoma}) 
can be rewritten by taking the oscillation average as
\begin{align}
\dot{J}
+ c \vev{H}J
&= 0,
\label{eq:LHdifeq}
\end{align}
where $c$ is $\mathcal O(1)$ numerical coefficient which can be determined explicitly if we specify the model.
Here we used the fact that $J$ is an adiabatic invariant, and thus
$\vev{HJ} = \vev{H}J$. Also remember that $J$ is proportional to $\mathcal{L}_{H}$.
Eq.~(\ref{eq:LHdifeq}) can be used to express $J$ as a function of the scale factor $a$. Since we know the relations
among $\vev{H\mathcal{L}_{H}}$, $\vev{\phi\mathcal{L}_{\phi}}$ and $\langle\dot{\phi}\mathcal{L}_{\phi} \rangle$, 
in general we can express $J$ only by $\vev{H}$, {\it i.e.}~$J \propto \vev{H}^{m}$, where $m$ is some number.
Then, the scale factor dependence of $\vev{H}$ and hence the expansion law of the universe can be estimated.
We will see how this procedure works for some specific models in the next section.

%%%%%%%%%%%%%%%%%%%%%%%%%%%%%%%%%%%%%%%%%%%%%%%%%%
\subsection{Particle production}
%%%%%%%%%%%%%%%%%%%%%%%%%%%%%%%%%%%%%%%%%%%%%%%%%%
As we said repeatedly, the Hubble parameter can violently oscillate with time for the general
Lagrangian~\eqref{eq:gravphiaction}. In such a case, the violent oscillation of the background
metric causes production of non-Weyl invariant particles~\cite{Ema:2015dka}.
Here, we briefly summarize basic ingredients of such a particle production event. More detailed
discussion is given in~\cite{Ema:2015oaa}

We call the produced (scalar) particle $\chi$ here. The parameter, $q$,  is defined as
\begin{align}
q \equiv \frac{|\Delta m_{\chi}^{2}|}{m_{\rm{eff}}^{2}},
\end{align}
where we have assumed that the $\chi$'s mass squared 
has an oscillating part $\Delta m_{\chi}^{2}$ due to the inflaton oscillation.
Then, the decay rate of $\phi$ is estimated as
\begin{align}
\Gamma_{\phi\rightarrow\chi} \sim \frac{q^{2}m_{\rm{eff}}^{3}}{\Phi_{c}^{2}},
\end{align}
where $\Phi_{c}$ is the canonically normalized amplitude of $\phi$. In the case where $\chi$ is minimally coupled
to gravity and the bare mass of $\chi$ is negligible, the effective mass of 
canonically normalized $\chi$ is given by the Hubble parameter squared and its time derivative
since the normalization of $\chi$ depends on the scale factor.
%$\chi$ is just the usual Hubble-induced mass,
Hence, the decay rate is given by
\begin{align}
\Gamma_{\phi\rightarrow\chi} \sim \frac{\left(\delta \dot{H}\right)^{2}}{m_{\rm{eff}}\Phi_{c}^{2}},
\label{eq:Gamma}
\end{align}
where $\delta H$ is the oscillating part of the Hubble parameter. Here, we have used the fact 
that $\delta \dot{H} \gtrsim \delta (H^{2})$ is always satisfied in the inflaton oscillation regime.

Note that by using the fact that $J$ is an adiabatic invariant, the oscillating part of the Hubble 
parameter can easily be deduced from the relation between $J$ and $H$. This is another benefit
of the adiabatic invariant. We will see this point for some specific models in the next section.

%%%%%%%%%%%%%%%%%%%%%%%%%%%%%%%%%%%%%%%%%%%%%%%%%%
\section{Examples}
\label{sec:examples}
\setcounter{equation}{0}
%%%%%%%%%%%%%%%%%%%%%%%%%%%%%%%%%%%%%%%%%%%%%%%%%%

In this section, we show several examples to illustrate the procedure explained in Sec.~\ref{sec:general}.
We assume that the inflaton potential (the $X$-independent part of $-G_{2}$) has the following monomial form\footnote{
Even for a more general form of potential, we can approximate the potential as the dominant part of it.
Then, the situation typically reduces to the monomial potential~\eqref{eq:pot}.
}
\begin{align}
V(\phi)
&= \frac{\lambda}{n}\phi^n,
\label{eq:pot}
\end{align}
and that the inflaton is oscillating around the minimum of the potential with amplitude $\Phi$.

%%%%%%%%%%%%%%%%%%%%%%%%%%%%%%%%%%%%%%%%%%%%%%%%%%
\subsection{Example 1}
\label{subsec:example1}
%%%%%%%%%%%%%%%%%%%%%%%%%%%%%%%%%%%%%%%%%%%%%%%%%%
We first consider
\begin{align}
G_2
&= X - V(\phi), \\
G_4
&= f(\phi),
\end{align}
where $f(\phi)$ is assumed to be of the form
\begin{align}
f(\phi)
= \frac{1}{2}M_P^2 \left(1 + f_1 \frac{\phi}{M_P}\right).
\end{align}
Here $\phi$ is understood as the deviation from the potential minimum\footnote{
Strictly speaking, $\phi = 0$ does not correspond to the minimum energy for the homogeneous $\phi$-condensation,
around which $\phi$ oscillates,
since the stationary point of $\phi$ receives contribution from $f_1\phi/M_P$ term. 
However we neglect such a small effect here.
}. 
In this model $\phi$ starts oscillating around $\phi \sim M_P / f_1$, therefore we assume $|f_1 \phi |/M_P \ll 1$.

%%%%%%%%%%%%%%%%%%%%%%%%%%%%%%%%%%%%%%%%%%%%%%%%%%
\subsubsection{Adiabatic invariant}
%%%%%%%%%%%%%%%%%%%%%%%%%%%%%%%%%%%%%%%%%%%%%%%%%%

The background action is given by
\begin{align}
S_G
&= \int d^4x \; a^3
\left[ -3M_P^2H^2 \left( 1 + f_1\frac{\phi}{M_P} \right) - 3M_P^2Hf_1\frac{\dot{\phi}}{M_P}
+ \frac{1}{2}\dot{\phi}^2 - V\right]
\end{align}
after integration by parts. Thus, $\mathcal{L}_{H}$ is given by
\begin{align}
{\mathcal L}_H
&= -6M_P^2H\left(1 + f_1\frac{\phi}{M_P} \right) - 3M_P^2f_1\frac{\dot{\phi}}{M_P},
\label{eq:LH_ex1}
\end{align}
and hence the adiabatic invariant $J$ is 
\begin{align}
J = H\left(1+f_{1}\frac{\phi}{M_{P}}\right) + \frac{f_{1}}{2}\frac{\dot{\phi}}{M_{P}}. 
\label{eq:J_ex1}
%\\
%\rho_J
%&\simeq \left( 1 + f_1\frac{\phi}{M_P} \right)
%\left( \frac{1}{2}\dot{\phi}^2 + V \right) 
%+ \frac{3}{4} f_1^2 \dot{\phi}^2,
\end{align}
%where we have used the Friedmann equation to obtain the expression of $\rho_{J}$.

%%%%%%%%%%%%%%%%%%%%%%%%%%%%%%%%%%%%%%%%%%%%%%%%%%
\subsubsection{Expansion law}
%%%%%%%%%%%%%%%%%%%%%%%%%%%%%%%%%%%%%%%%%%%%%%%%%%

\begin{figure}
\begin{minipage}{0.5\hsize}
\begin{center}
\includegraphics[scale=0.85]{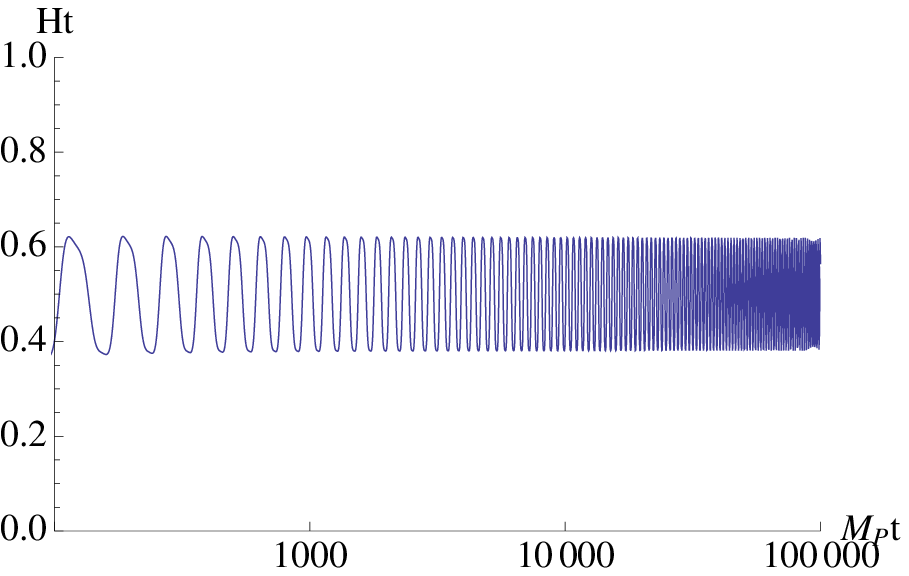}
\end{center}
\end{minipage}
\begin{minipage}{0.5\hsize}
\begin{center}
\includegraphics[scale=0.85]{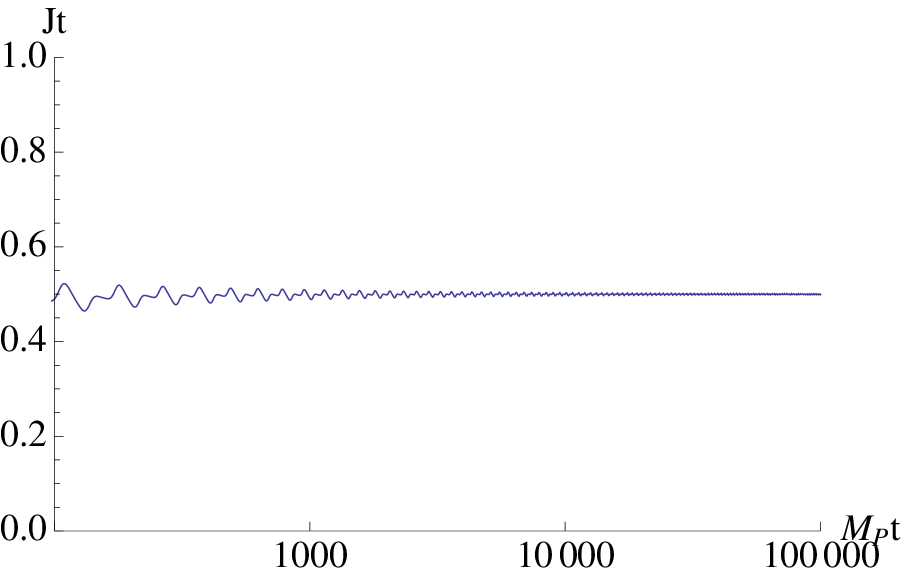}
\end{center}
\end{minipage}
\caption {\small
Time evolution of $H$ and $J$ for Example 1 is shown. 
In these figures, we have taken
$V = \lambda \phi^4/4$, $\lambda = 1$ and $f_1 = 0.2$. 
$H$ oscillates violently ({\it Left}) while $J$ does not ({\it Right}).
}
\label{fig:ex1}
\end{figure}

Since we assumed $| f_1\phi | / M_P \ll 1$, the Lagrangian is approximated as
\begin{align}
{\mathcal L}
&\simeq -3M_P^2H^2 - 3M_P^2Hf_1\frac{\dot{\phi}}{M_P} + \frac{1}{2}\dot{\phi}^2 - V.
\label{eq:L_ex1_dom}
\end{align}
Thus, it can be decomposed as
\begin{align}
{\mathcal L}
&\simeq \frac{1}{2}H{\mathcal L}_H + \frac{1}{2}\dot{\phi}{\mathcal L}_{\dot{\phi}} + \frac{1}{n}\phi{\mathcal L}_\phi.
\label{eq:L_ex1}
\end{align}
Note that this equation is the same as the one in the minimal case of Einstein gravity, where the second term in Eq.~(\ref{eq:L_ex1_dom}) does not exist.
Eqs.~(\ref{eq:eomconstave}) and (\ref{eq:virial}) together with the decomposition Eq.~(\ref{eq:L_ex1}) gives
\begin{align}
\vev{{\mathcal L}}
&= \frac{2}{n+2}\vev{H{\mathcal L}_H}.
\end{align}
Substituting this into oscillation-averaged Eq.~(\ref{eq:eoma}) gives
\begin{align}
\dot{J} + \frac{3n}{n+2}\vev{H}J
&= 0,
\end{align}
and hence we obtain
\begin{align}
J \propto a^{-\frac{3n}{n+2}}.
\end{align}
Note that $\vev{\dot{\phi}} = 0$, and thus the expansion law of the universe is estimated as
\begin{align}
\vev{H} \simeq J
&\propto a^{-\frac{3n}{n+2}},
\end{align}
or
\begin{align}
\vev{H}
&\simeq \frac{n+2}{3n}\frac{1}{t}.
\end{align}
This expansion law is the same as that in the minimal case of Einstein gravity.
We plot the time dependence of $H$ and $J$ (multiplied by $t$) for $n=4$ in Fig.~\ref{fig:ex1}.

%%%%%%%%%%%%%%%%
%%%%%%%%%%%%%%%%

%%%%%%%%%%%%%%%%%%%%%%%%%%%%%%%%%%%%%%%%%%%%%%%%%%
\subsubsection{Particle production}
%%%%%%%%%%%%%%%%%%%%%%%%%%%%%%%%%%%%%%%%%%%%%%%%%%

Recalling that J is conserved during the inflaton oscillation, 
the oscillating part of the Hubble parameter is deduced from Eq.~\eqref{eq:J_ex1} as
\begin{align}
\delta H
&\simeq -f_1\frac{\dot{\phi}}{2M_P}.
\end{align}
Then, from Eq.~(\ref{eq:Gamma}), the production rate is estimated as
\begin{align}
\Gamma_{\phi \to \chi}
&\sim f_{1}^{2}\frac{m_{\rm{eff}}^{3}}{M_{P}^{2}},
\end{align}
where $m_{\rm eff} \sim \lambda^{1/2} \Phi^{n/2-1}$.
This reproduces the result in \cite{Watanabe:2006ku,Ema:2015dka}.

%%%%%%%%%%%%%%%%%%%%%%%%%%%%%%%%%%%%%%%%%%%%%%%%%%
\subsection{Example 2}
\label{subsec:example2}
%%%%%%%%%%%%%%%%%%%%%%%%%%%%%%%%%%%%%%%%%%%%%%%%%%
As a second example, we consider
\begin{align}
G_2
&= X - V(\phi), \\
G_4
&= \frac{1}{2}M_P^2 + \frac{X}{2M^2}.
\end{align}
Note that this system is equivalent to 
\begin{align}
G_2
&= X - V(\phi), \\
G_4
&= \frac{1}{2}M_P^2, \\
G_5
&= -\frac{\phi}{2M^2},
\end{align}
where $M_P$ is the reduced Planck mass and $M$ is some mass parameter.

%%%%%%%%%%%%%%%%%%%%%%%%%%%%%%%%%%%%%%%%%%%%%%%%%%
\subsubsection{Adiabatic invariant}
%%%%%%%%%%%%%%%%%%%%%%%%%%%%%%%%%%%%%%%%%%%%%%%%%%
The background action becomes
\begin{align}
S_G
&= \int d^4x \; a^3
\left[-3M_P^2H^2 + \left( 1 + \frac{3H^2}{M^2} \right)\frac{\dot{\phi}^2}{2} - V \right],
\end{align}
after integration by parts. Therefore, $\mathcal{L}_{H}$ is given by
\begin{align}
\mathcal{L}_{H} = -6M_{P}^{2}H\left(1-\frac{\dot{\phi}^{2}}{2M_{P}^{2}M^{2}}\right),
\end{align}
and the adiabatic invariant $J$ is given by
\begin{align}
J &= H\left(1-\frac{\dot{\phi}^{2}}{2M_{P}^{2}M^{2}}\right).
\label{eq:J_ex2}
\end{align}
%%

%%%%%%%%%%%%%%%%%%%%%%%%%%%%%%%%%%%%%%%%%%%%%%%%%%
\subsubsection{Expansion law}
\begin{figure}
\begin{minipage}{0.5\hsize}
\begin{center}
\includegraphics[scale=0.85]{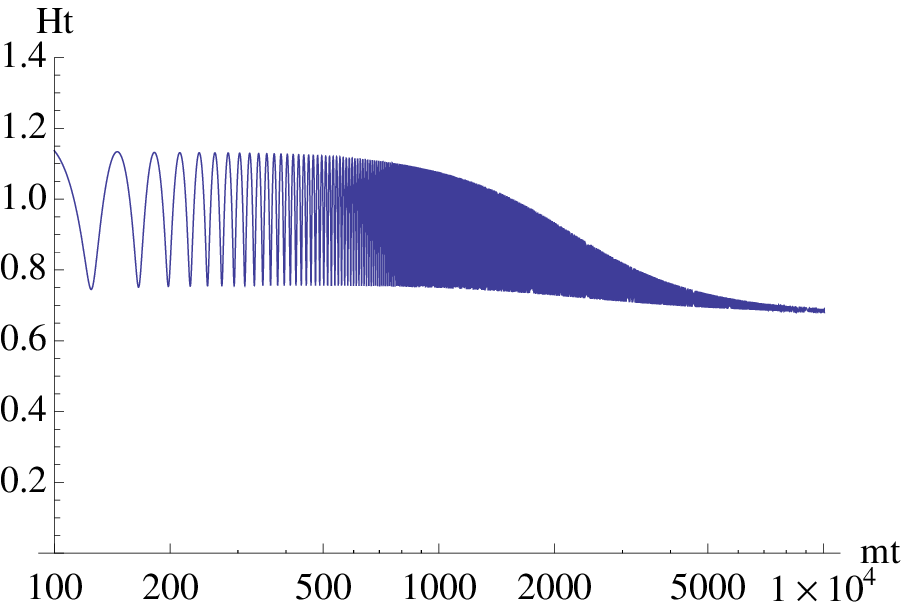}
\end{center}
\end{minipage}
\begin{minipage}{0.5\hsize}
\begin{center}
\includegraphics[scale=0.85]{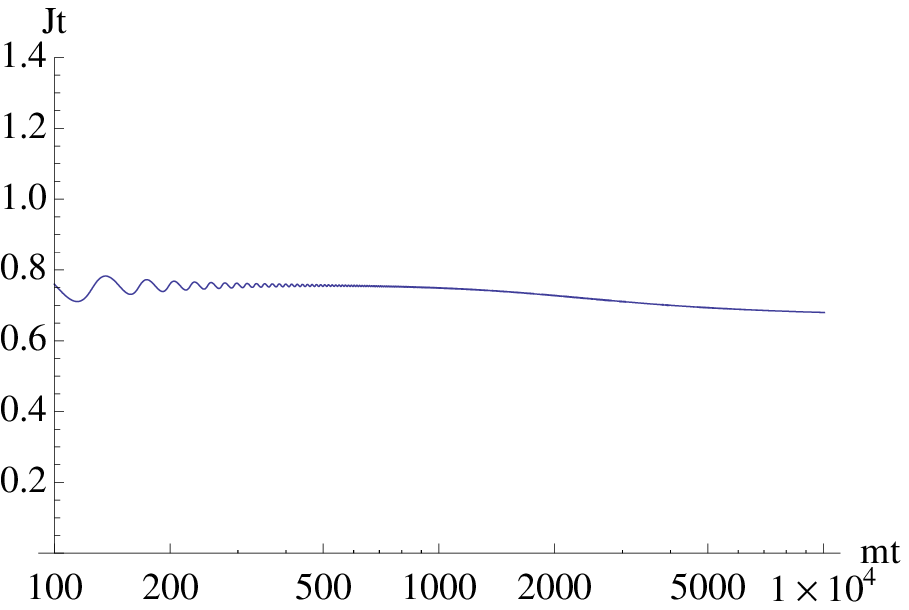}
\end{center}
\end{minipage}
\caption {\small
Time evolution of $H$ and $J$ for Example 2 is shown.
In these figures, we have taken
$V=m^2\phi^2/2$ and $m/M = 10^3$.
{\it Left}: Plot of $Ht$. $H$ violently oscillates in the non-minimal regime $H^2/M^2 \gg 1$. 
The expansion law is $\vev{H}t = 1$ for this regime, 
while it gradually approaches to the value with canonical kinetic term $2/3$ for $H^2/M^2 \ll 1$.
{\it Right}: Plot of $Jt$. It has a suppressed amplitude of oscillation 
compared to $Ht$.
}
\label{fig:ex2}
\end{figure}

%%%%%%%%%%%%%%%%%%%%%%%%%%%%%%%%%%%%%%%%%%%%%%%%%%
The background dynamics of this system is extensively studied in \cite{Jinno:2013fka,Ema:2015oaa}, 
where it is shown that this system undergoes a nontrivial regime when $H^2/M^2 \gg 1$.
Let us focus on such a regime, when the Lagrangian is approximated as
\begin{align}
{\mathcal L}
&\simeq -3M_P^2H^2 + \frac{3H^2}{M^2}\frac{\dot{\phi}^2}{2} - V.
\end{align}
Then, it can be decomposed as
\begin{align}
{\mathcal L}
&\simeq \frac{1}{2}H{\mathcal L}_H + \frac{1}{n}\phi{\mathcal L}_\phi.
\label{eq:Ldoc_ex2}
\end{align}
Eqs.~(\ref{eq:eomconstave}), (\ref{eq:virial}) and (\ref{eq:Ldoc_ex2}) give
\begin{align}
\vev{{\mathcal L}}
&= \frac{n+2}{2(n+1)}\vev{H{\mathcal L}_H}.
\end{align}
Substituting this into oscillation-averaged Eq.~(\ref{eq:eoma}), one obtains the evolution equation for 
the adiabatic invariant $J$:
\begin{align}
\dot{J}
+ \frac{3n}{2(n+1)} \vev{H} J 
= 0.
\end{align}
From this equation one can derive 
\begin{align}
J
&\propto a^{-\frac{3n}{2(n+1)}}.
\end{align}
Now let us estimate the expansion law of the universe.
Since $J \propto H(1-\dot{\phi}^2/2M_P^2M^2)$,
and the amplitude of $\dot{\phi}^2 / M_P^2M^2$ is constant during the regime we consider,\footnote{
This can be seen by the Friedmann equation.
See \cite{Jinno:2013fka,Ema:2015oaa} for detail.
}
we have
\begin{align}
\vev{H}
\propto a^{-\frac{3n}{2(n+1)}},
\end{align}
and hence the expansion law is given by
\begin{align}
\vev{H}
&= \frac{2(n+1)}{3n}\frac{1}{t}.
\end{align}
We plot the time dependence of $H$ and $J$ (multiplied by $t$) for $n = 2$ in Fig.~\ref{fig:ex2}\footnote{
The center of the oscillation in the left panel of Fig.~\ref{fig:ex2} is not $1$, 
but time-averaging of $H$ gives $\vev{H}t = 1$.
}.

%%%%%%%%%%%%%%%%
%%%%%%%%%%%%%%%%

%%%%%%%%%%%%%%%%%%%%%%%%%%%%%%%%%%%%%%%%%%%%%%%%%%
\subsubsection{Particle production}
%%%%%%%%%%%%%%%%%%%%%%%%%%%%%%%%%%%%%%%%%%%%%%%%%%

From Eq.~\eqref{eq:J_ex2}, the oscillating part of the Hubble parameter is deduced as
\begin{align}
\delta H &\sim \frac{J\dot{\phi}^{2}}{M_{P}^{2}M^{2}}.
\end{align}
Thus, the decay rate of $\phi$ is estimated as
\begin{align}
\Gamma_{\phi \to \chi}
\sim \frac{m_{\rm eff}^5\Phi^2}{M_P^4M^2},
\end{align}
where we have taken $J \sim H$ as an order estimation, and $\Phi_{c} \sim H\Phi/M$. 
Also, $m_{\rm eff} \sim (M/H) \lambda^{1/2} \Phi^{n/2-1}$.
This reproduces the result shown in \cite{Ema:2015oaa} 
(for case B, where the non-minimal coupling effect is larger than the one that exists in the minimal setup).\footnote{
In fact, from the Friedmann equation, one can show that $m_{\rm{eff}}^{2}\Phi^{2}/M_{P}^{2}M^{2} \sim \mathcal{O}(1)$ is 
always satisfied for $H \gg M$ in this model. Then, the decay rate reduces to 
$\Gamma_{\phi\to\chi} \sim M^{2}m_{\rm{eff}}/\Phi^{2}$, which is the same as that obtained in \cite{Ema:2015oaa}.
}

%%%%%%%%%%%%%%%%%%%%%%%%%%%%%%%%%%%%%%%%%%%%%%%%%%
\subsection{Example 3}
\label{subsec_example3}
%%%%%%%%%%%%%%%%%%%%%%%%%%%%%%%%%%%%%%%%%%%%%%%%%%
Finally let us consider the system
\begin{align}
G_2
&= X - V(\phi), \\
G_4
&= \frac{1}{2}M_P^2, \\
G_5
&= -\frac{\phi^3}{M^4},
\end{align}
where $M$ is some mass parameter.

%%%%%%%%%%%%%%%%%%%%%%%%%%%%%%%%%%%%%%%%%%%%%%%%%%
\subsubsection{Adiabatic invariant}
%%%%%%%%%%%%%%%%%%%%%%%%%%%%%%%%%%%%%%%%%%%%%%%%%%

The background action is 
\begin{align}
S_{G}
&= \int d^4x \; a^3 
\left[ -3M_P^2H^2 + \left( 1 + \frac{18H^2\phi^2}{M^4} \right)\frac{\dot{\phi}^2}{2} - V \right],
\end{align}
from which $\mathcal{L}_{H}$ is given by
\begin{align}
{\mathcal L}_H
&= -6M_P^2 H\left( 1 - \frac{3\phi^2\dot{\phi}^2}{M_P^2M^4} \right).
\end{align}
Then, the adiabatic invariant $J$ is given by
\begin{align}
J &= H\left(1-\frac{3\phi^{2}\dot{\phi}^{2}}{M_{P}^{2}M^{4}}\right). \label{eq:J_ex3}
\end{align}
%%

%%%%%%%%%%%%%%%%%%%%%%%%%%%%%%%%%%%%%%%%%%%%%%%%%%
\subsubsection{Expansion law}
\begin{figure}
\begin{minipage}{0.5\hsize}
\begin{center}
\includegraphics[scale=0.85]{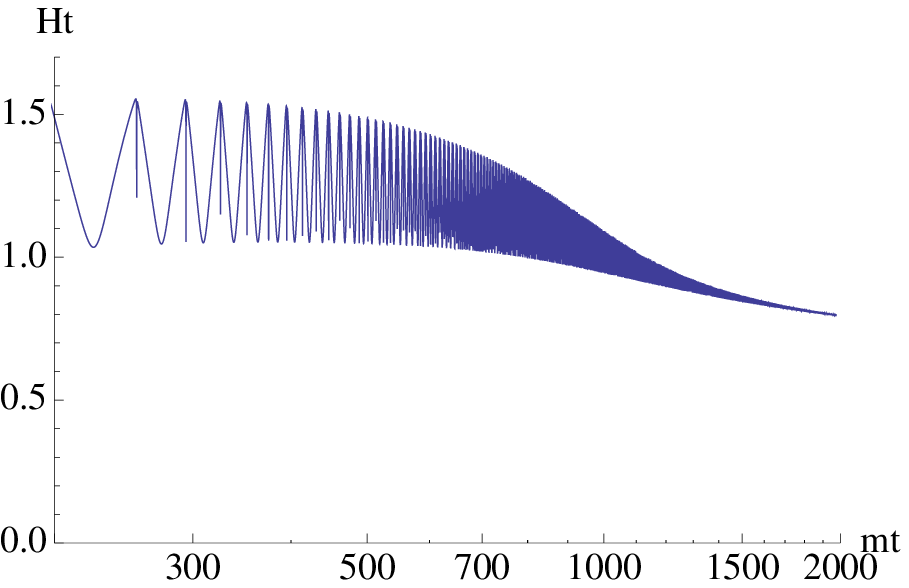}
\end{center}
\end{minipage}
\begin{minipage}{0.5\hsize}
\begin{center}
\includegraphics[scale=0.85]{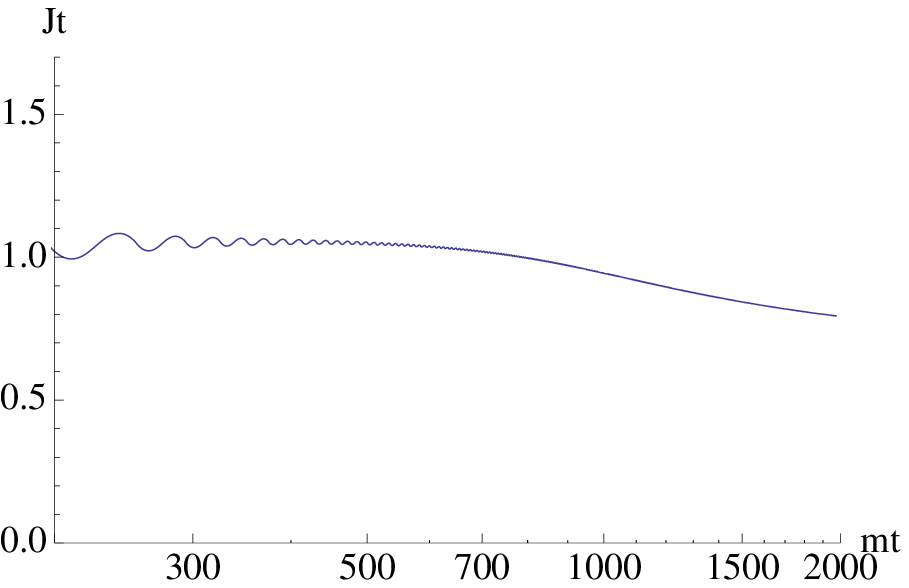}
\end{center}
\end{minipage}
\caption {\small
Time evolution of $H$ and $J$ for Example 3 is shown. 
In these figures, we have taken $V=m^2\phi^2/2$, $m/M_P = 1$ and $M/M_P = 10^{-5/2}$. 
{\it Left}: Plot of $Ht$. $H$ violently oscillates in the non-minimal regime $H^2\phi^2/M^4 \gg 1$. 
The expansion law is $\vev{H}t = 4/3$ for this regime, 
while it gradually approaches to the value with canonical kinetic term $2/3$ for $H^2\phi^2/M^4 \ll 1$.
{\it Right}: Plot of $Jt$. It has a suppressed amplitude of oscillation 
compared to $Ht$.
}
\label{fig:ex3}
\end{figure}

%%%%%%%%%%%%%%%%%%%%%%%%%%%%%%%%%%%%%%%%%%%%%%%%%%
We focus on the regime when the non-minimal kinetic term takes a dominant role here. Then,
the Lagrangian is approximated as
\begin{align}
{\mathcal L}
&\simeq -3M_P^2H^2 + \frac{18H^2\phi^2}{M^4}\frac{\dot{\phi}^2}{2} - V,
\end{align}
and hence it is decomposed as
\begin{align}
{\mathcal L}
&\simeq \frac{1}{2}H{\mathcal L}_H + \frac{1}{n}\phi{\mathcal L}_\phi - \frac{1}{n}\dot{\phi}{\mathcal L}_{\dot{\phi}}.
\label{eq:3rdrel_ex3}
\end{align}
Therefore, Eqs.~(\ref{eq:virial}), (\ref{eq:eomconstave}) and (\ref{eq:3rdrel_ex3}) give
\begin{align}
\vev{\mathcal L}
&= \frac{n+4}{2(n+2)}\vev{H{\mathcal L}_H}.
\end{align}
Substituting this into Eq.~\eqref{eq:eoma}, we get
\begin{align}
\dot{J} 
+ \frac{3n}{2(n+2)}\vev{H}J
= 0,
\end{align}
and thus
\begin{align}
J
&\propto a^{-\frac{3n}{2(n+2)}}.
\end{align}
Since the absolute value of the second term in the parenthesis in Eq.~(\ref{eq:J_ex3}) is less than $1/3$, 
as is guaranteed by the Friedmann equation, this term does not affect the power-law form of $J$.
Thus,
\begin{align}
\vev{H}
&\propto a^{-\frac{3n}{2(n+2)}}
\end{align}
and the expansion law of the universe is given by
\begin{align}
\vev{H}
&= \frac{2(n+2)}{3n}\frac{1}{t}.
\end{align}
We plot $H$ and $J$ (multiplied by $t$) for $n=2$ in Fig.~\ref{fig:ex3}\footnote{
As seen in Fig.~\ref{fig:ex3}, there are spikes in the Hubble parameter.
These correspond to $\phi \simeq 0$, 
and their typical time scale is $\sim m_{\rm eff}/m^2\,(\ll 1/m_{\rm eff})$. 
At these spikes the quantity in the parenthesis in the RHS of Eq.~(\ref{eq:J_ex3}) 
changes from $2/3$ to $1$ and thus $H$ changes from the maximum to the minimum of its oscillation. 
}.

%%%%%%%%%%%%%%%%
%%%%%%%%%%%%%%%%

%%%%%%%%%%%%%%%%%%%%%%%%%%%%%%%%%%%%%%%%%%%%%%%%%%
\subsubsection{Particle production}
%%%%%%%%%%%%%%%%%%%%%%%%%%%%%%%%%%%%%%%%%%%%%%%%%%
From Eq.~\eqref{eq:J_ex3}, the oscillating part of the Hubble parameter is
\begin{align}
\delta H \sim J\frac{\phi^{2}\dot{\phi}^{2}}{M_{P}^{2}M^{4}},
\end{align}
and hence the decay rate is estimated as
\begin{align}
\Gamma_{\phi\to\chi} \sim \frac{m_{\rm{eff}}^{5}\Phi^{4}}{M_{P}^{4}M^{4}},
\end{align}
where we have taken $J\sim H$ as an order estimation, and $\Phi_{c}\sim H\Phi^{2}/M^{2}$.\footnote{
This production rate may be enhanced by the effect of spikes mentioned in the previous footnote.
}
Also, $m_{\rm eff} \sim (M^2 / H\Phi)\lambda^{1/2}\Phi^{n/2-1}$.

%%%%%%%%%%%%%%%%%%%%%%%%%%%%%%%%%%%%%%%%%%%%%%%%%%
\section{Conclusion}
\label{sec_conclusion}
\setcounter{equation}{0}
%%%%%%%%%%%%%%%%%%%%%%%%%%%%%%%%%%%%%%%%%%%%%%%%%%

In this paper, we study the generalized Galileon theories in the context of inflation, 
especially focusing on the inflaton oscillation regime. During this regime,
in general, these models lead to violent oscillations of the Hubble parameter and the
energy density of inflaton, and hence it would be desirable to construct an adiabatic quantity
which evolves only with the time scale of $H^{-1}$.
We have shown the existence of an adiabatic invariant $\dot{Q} = {\mathcal O}(HQ)$ 
during this regime,
and also shown an explicit method to construct this adiabatic invariant.
To confirm our procedure,
we have demonstrated that the proposed adiabatic invariant 
actually evolves with the time scale of $H^{-1}$ in several examples of generalized Galileon theories 
by numerical calculation.
Such an invariant is useful in estimating the expansion law of the universe,
and expressing the oscillation in the Hubble parameter in terms of the inflaton so as to 
estimate the amount of particle production via gravitational effects.
Thus it may take crucial roles to predict the precise value of the spectral index.

We also emphasize that the analysis background dynamics is a first step to understand the actual phenomena
in the oscillation regime.
In a specific example of $G^{\mu\nu}\partial_\mu\phi\partial_\nu\phi$ model, the relation $|\dot H| \sim m_{\rm eff} H \gg H^2$
was crucial for the Laplacian instability~\cite{Ema:2015oaa}.
Without correct understandings of the background dynamics, we cannot judge the phenomenological validity of various inflation models.

%%%%%%%%%%%%%%%%%%%%%%%%%%%%%%%%%%%%%%%%%%%%
\section*{Acknowledgments}
%%%%%%%%%%%%%%%%%%%%%%%%%%%%%%%%%%%%%%%%%%%%

This work was supported by the Grant-in-Aid for Scientific Research on Scientific Research A (No.26247042 [KN]),
Young Scientists B (No.26800121 [KN]) and Innovative Areas (No.26104009 [KN]).
This work was supported by World Premier International Research Center Initiative (WPI Initiative), MEXT, Japan. 
The work of Y.E., R.J. and K.M. was supported in part by JSPS Research Fellowships for Young Scientists.
The work of Y.E. and R.J. was also supported in part by the Program for Leading Graduate Schools, MEXT, Japan.

%%%%%%%%%%%%%%%%%%%%%%%%%%%%%%%%%%%%%%%%%%%%%%%%%%

%%%%%%%%%%%%%%%%%%%%%%%%%%%%%%%%%%%%%%%%%%%%%%%%%%

\end{document}